\documentclass[%
 aip,
 apl,
 amsmath,amssymb,
 reprint,%
floatfix]{revtex4-1}

\usepackage{graphicx}
\usepackage{subcaption}
\usepackage{dcolumn}
\usepackage{bm}

\usepackage[utf8]{inputenc}
\usepackage[T1]{fontenc}
\usepackage{mathptmx}

\begin{document}
\captionsetup{justification=centerlast}

\preprint{AIP/123-QED}

\title[Terahertz continuum generation...]{Terahertz continuum generation in the LCS
lattice}

\author{Qiutong Jin}
 \email{qiutong-jin@uiowa.edu.}
\author{David R. Andersen}%
 \email{k0rx@uiowa.edu.}
 \altaffiliation[Also at ]{Department of Physics and Astronomy, University of Iowa.}
\affiliation{
Electrical and Computer Engineering, University of Iowa, Iowa City, IA 52242,
USA }%

\date{\today}

\begin{abstract}
Rabi oscillations in two-level Dirac systems have been shown to
alter the frequency content of the system's nonlinear response.
In particular, when considering Rabi oscillations in a quantum model
beyond the semiclassical Boltzmann theory, even harmonics may
be generated despite the centrosymmetric nature of these systems.
This effect magnifies with increasing excitation intensity.
In this work, we extend the Rabi theory to a
three-level Dirac system arising from a line-centered-square
optical lattice. In this case, the Dirac cones are bisected
at the Dirac point by a flat band that persists throughout the
Brillioun zone. Due to the presence of this flat band, we expect a
significant enhancement of the coupling between Dirac states,
resulting in a large increase of the Rabi effects and the
associated nonlinearities, leading to continuum generation of
terahertz radiation.
\end{abstract}

\maketitle

\section{\label{sec:level1}INTRODUCTION\label{sec:introduction}}
Due to the novel electronic \cite{castro-neto}, thermal
\cite{thermal-nanoribbon}, mechanical \cite{mechanical} and optoelectronic
\cite{finestructure} properties it possesses, graphene
has drawn a considerable
amount of attention in many fields. Its Dirac band structure
contributes to the optical response of graphene, giving rise to novel effects
including efficient light absorption and a large nonlinear
susceptibility. The harmonic
response of the current density in graphene, based on the semiclassical Boltzmann
theory and the centrosymmetric structure of graphene itself, was believed to be
restricted to only odd harmonic spectra. However, Lee \textit{et. al.}
\cite{rabi} demonstrated that even in the presence of the centrosymmetry, the
nonlinear optical response in graphene is not limited to odd harmonic spectra
when the electron dynamics of Rabi oscillations are included in the current response
\cite{rabinlo}. The Rabi oscillation can shift the current-induced harmonic spectra of graphene and thus lead to a peak at the spectral position of the second harmonic of the incident light.

\begin{figure}[tph]
  \centering
\begin{subfigure}[b]{0.2\textwidth}
  \includegraphics[width=2.8cm, height=2.8cm]{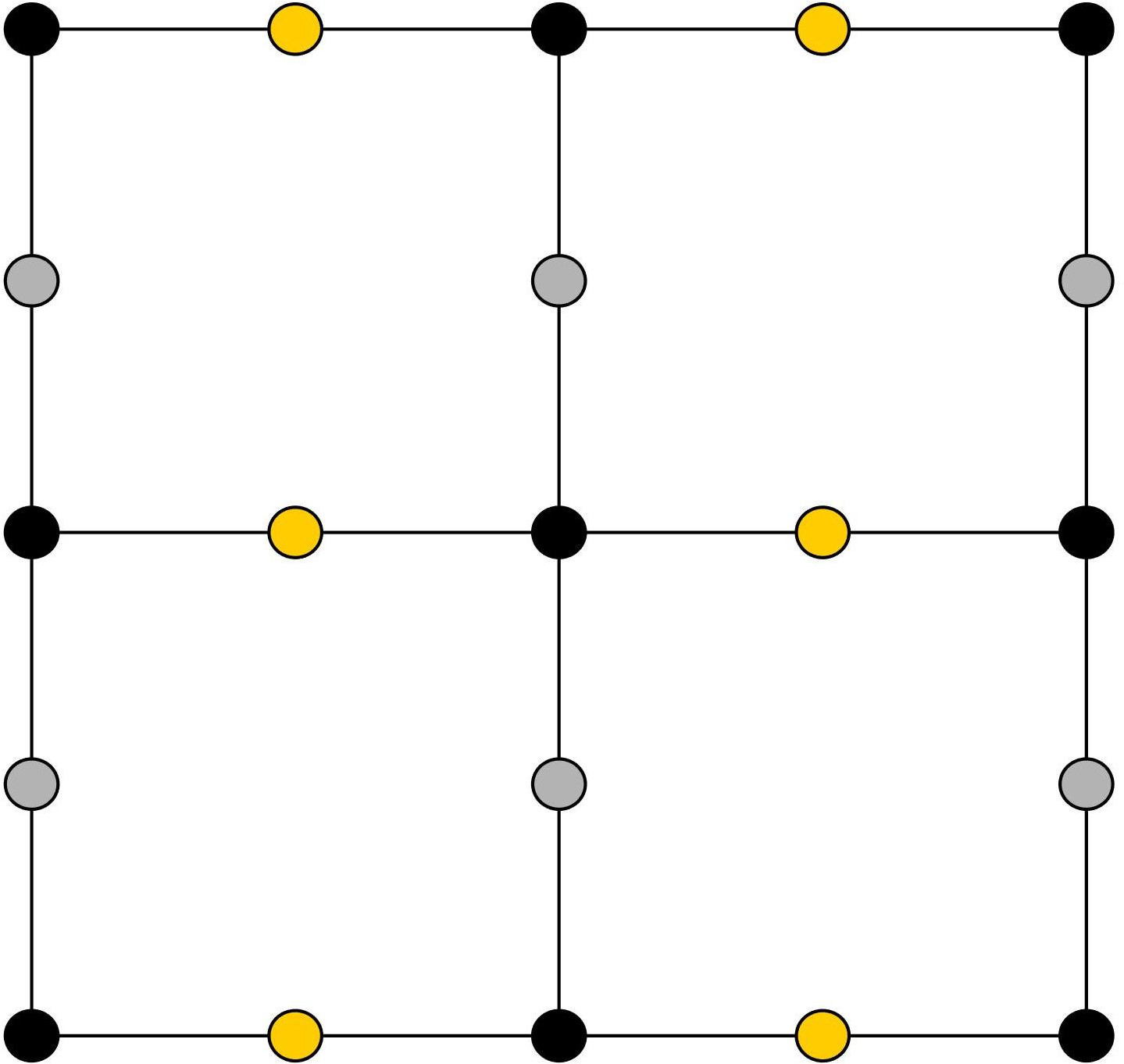}
  \caption{}
  \label{fig:lcslattice}
\end{subfigure}
\quad
\begin{subfigure}[b]{0.2\textwidth}
  \includegraphics[width=1.8cm,height=2.8cm]{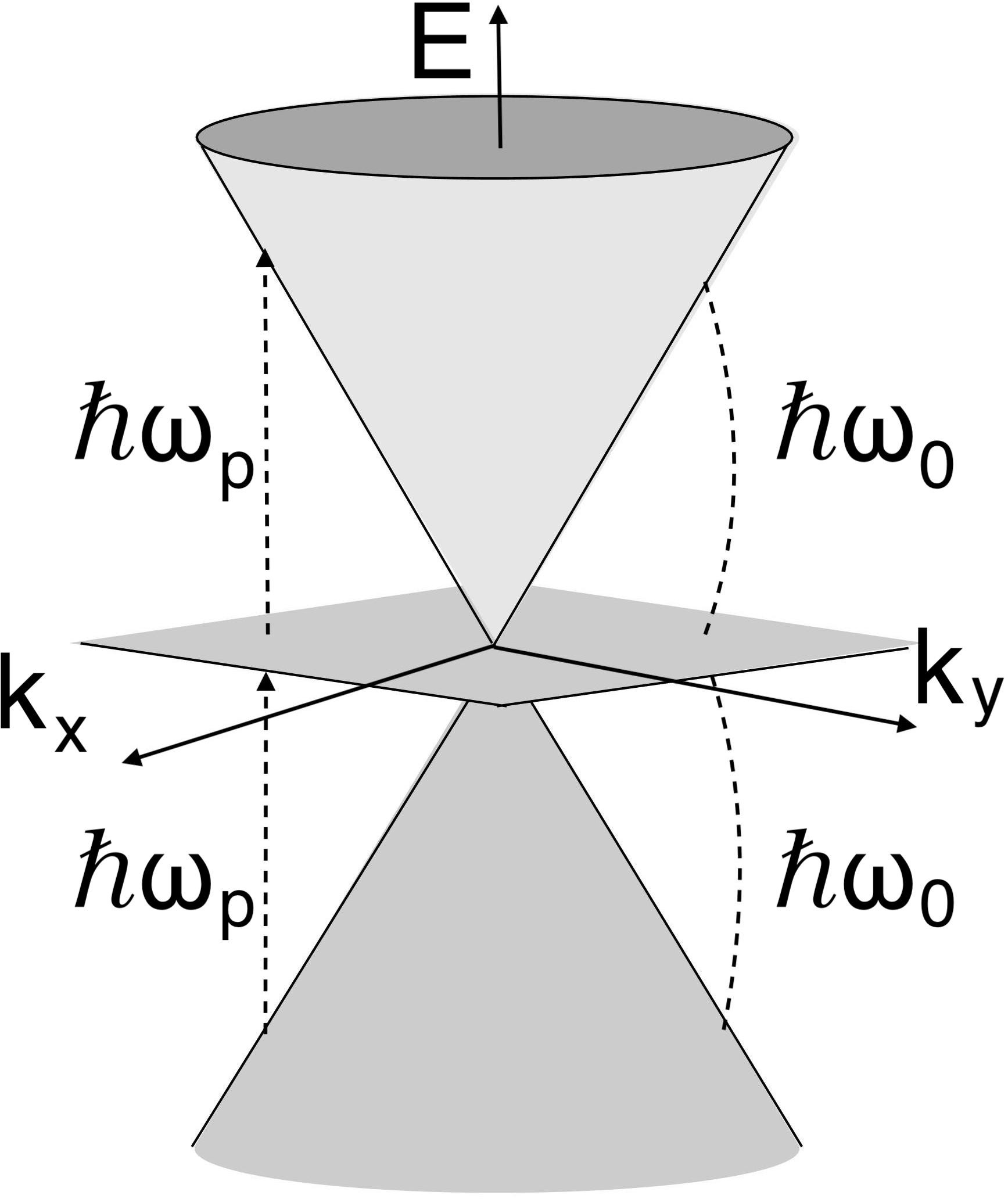}
  \caption{}
  \label{fig:lcsbandstructure}
\end{subfigure}
\vspace{-0.1in}
\caption{Schematic illustration of the LCS lattice (a) configuration and (b) energy band
structure.}
\label{fig:latticeandbandstructure}
\end{figure}

The line-centered-square (LCS) lattice (Fig. \ref{fig:lcslattice}), formed by using ultracold atoms
trapped in an optical lattice, was first described by Shen \textit{et. al.}
\cite{lcs}. It is a two-dimensional counterpart of the face-centered-cubic
lattice, and contains a single Dirac cone in the energy spectrum with an
additional flat band energy state touching at the Dirac point. Thus,
the LCS lattice exhibits three bands,
\textit{i.e.} the conduction band, the valence band and a central flat band
(Fig. \ref{fig:lcsbandstructure}).  The band structure of the LCS lattice
satisfies a three-component quantum equation for pseudospin 1 Fermions.

In this paper, we calculate the Rabi contribution to the harmonic spectra of
massless Dirac fermions in LCS lattice for various parameter sets. Following
Lee \textit{et. al.} \cite{rabi}, we begin with the time-dependent Dirac
equation (TDDE) to model LCS electron dynamics. By transforming the TDDE to the
dipole gauge (DG) \cite{dipolegauge}, we obtain the Rabi frequency of the LCS lattice, and
use it to calculate the Rabi contribution to the current response. We
demonstrate that, in analogy with graphene, the nonlinear optical response of
LCS lattice is not restricted to odd harmonic spectra when Rabi oscillation
contributes strongly to the current response. We find the first five orders of
the harmonic spectra, ranging from 2 THz to 10 THz, fuse into a continuous
spectrum in the presence of a 2 THz incident light field. Further,
odd harmonics up to $n=25$ exhibit significant energy content for reasonable
pump energy levels.

The paper is organized as follows. In Section \ref{sec:theoretical}, we begin
with the TDDE in the DG to obtain the Rabi frequency and current response
of the LCS lattice due to the terahertz (THz) light field. In Section \ref{sec:resultsanddiscussion}, we
elucidate the Rabi coupling effect on harmonic spectra. Finally, we draw
conclusions in Section \ref{sec:conclusion}.

\section{THEORETICAL MODEL\label{sec:theoretical}}


\subsection{Eigenstates of the unperturbed Hamiltonian}
In the LCS lattice, the Dirac Hamiltonian of a 2D massless
electron can be described \cite{lcs} as:
\begin{equation}
\mathbf{H}_0=\hbar v_0 \begin{pmatrix}
0 & k_x & 0 \\
k_x & 0 & k_y\\
0 & k_y & 0
\end{pmatrix}
\label{eq:unperturbedhamiltonian}
\end{equation}
where $v_0=\pm ta/\hbar$ is the Fermi velocity, $a$ is the LCS lattice constant,
$t$ is a
hopping parameter, $k_x$ and $k_y$ are the perturbation components of
${\bf k}$ from Dirac point, and $k = | \bf k |$. The eigenstates of
$\textit{H}_0$ are given by $\psi^{(i)}_{\bf k}({\bf r})=
(1/2\pi) \, \exp(\textnormal{i}{\bf k}\cdot{\bf r}) \, \mathbf{u}^{(i)}_{\bf k}$ with
eigenvalues 0, $-\hbar v_0 \sqrt[2]{{k_x}^2+{k_y}^2}$,and $\hbar v_0
\sqrt{{k_x}^2+{k_y}^2}$, and eigenvectors:
\begin{equation}
\mathbf{u}^{(1)}_{\bf k}=\begin{bmatrix}
-\frac{k_y}{k}\\ 0\\
\frac{k_x}{k}
\end{bmatrix},\ \mathbf{u}^{(2)}_{\bf k}=\begin{bmatrix}
\frac{k_x}{\sqrt{2}k}\\
-\frac{1}{\sqrt{2}}\\
\frac{k_y}{\sqrt{2}k}
\end{bmatrix},\  \mathrm{and}\ \mathbf{u}^{(3)}_{\bf k}=\begin{bmatrix}
\frac{k_x}{\sqrt{2}k}\\
\frac{1}{\sqrt{2}}\\
\frac{k_y}{\sqrt{2}k}
\end{bmatrix}
\label{eq:eigenvectors}
\end{equation}
The orthonormality of these eigenstates is given by $\int\textit{d}^2 {\bf
r}[\psi^{(s')}_{\bf k'}(\bf r)]^{\dagger}\psi^{(\textnormal{s})}_{\bf k}(\bf
r)=\delta ({\bf k}-{\bf k'})\delta_{ss'}$, with $s,s'=1,2,3$ indexing the pseudospin.
We also note here that the pseudospin indices 1, 2, and 3 correspond to the flat, valence,
and conduction bands respectively.

\subsection{TDDE in the dipole gauge with terahertz optical pump}
We define the incident terahertz optical pump field in the Coulomb gauge by the
vector potential:
\begin{equation}
{\bf A}(t)=\frac{E_0}{\omega_0}e^{-\frac{1}{2}{\ln 2}(\frac{t}{\tau})^2}[\hat{x }{ \cos (\frac{\zeta}{2})}{\cos(\omega_0 t)}+\hat{y }{ \sin(\frac{\zeta}{2})}{\sin(\omega_0 t)}],
\end{equation}
where $E_0$ is the peak of the incident electric-field strength, $\zeta$ is the polarization factor, $\omega_0=2\pi f_0$ is the central frequency and $\tau$ is the pulse width.
With this incident light field, the Dirac Hamiltonian may be written:
\begin{equation}
\hat{H}={\textit{v}_0}{\bf J}\cdot({{\hbar}{\bf\hat k}+q{\bf A}(t)}),
\end{equation}
where ${\bf J}=[{\bf {J_3}{S_-},{J_3}{S_+},0}]^\dagger$, ${\bf S_\pm}$ are the
spin-1 ladder operators, $\hbar \hat{\mathbf{k}}$ is the momentum operator, and
${\bf J_3}$ is the $3\times3$ exchange matrix. Using this Hamiltonian, we obtain
the TDDE in the Coulomb gauge:
\begin{equation}
i\hbar \, \frac{\partial \Psi({\bf r},t)}{\partial t}=\hat{H} \, \Psi({\bf r},t)
\label{eq:tdde}
\end{equation}

For a normally-incident THz optical field, the solution to Eq. \ref{eq:tdde} may be expressed as:
\begin{equation}
\Psi({\bf r},t)=\Psi^{D}({\bf r},t)\exp\left[\left(-i\left(\frac{e}{\hbar}\right)\bf r\cdot
A{\textnormal{(t)}}\right)\right]
\end{equation}
This represents a gauge transformation from Coulomb to dipole gauge where
$-i(\frac{e}{\hbar})\bf r\cdot A{\textnormal{(t)}}$ is the gauge generating
function for the transformation, and $\Psi^{D}({\bf r},t)$ is the spinor
wavefunction in the dipole gauge, which we expand in the eigenstates of
$\mathbf{H}_0$ as:
\begin{equation}
\Psi^{D}({\bf r},t)=c_{1}(t)\psi^{(1)}_{\bf k}({\bf r})+c_{2}(t)\psi^{(2)}_{\bf k}({\bf r})+c_{3}(t)\psi^{(3)}_{\bf k}({\bf r})
\label{eq:dipolepsi}
\end{equation}

Substituting $\Psi({\bf r},t)$ into Eq. \ref{eq:tdde} yields the TDDE in the
dipole gauge:
\begin{equation}
i\hbar \, \frac{\partial \Psi^{D}({\bf
r},t)}{\partial t}=[\mathbf{H}_0+\hat{V}(t)])\, \Psi^{D}({\bf r},t),
\label{eq:dipoletdde}
\end{equation}
with $\hat{V}(t)=e{\bf r}\cdot{\bf E}(t){\textit{I}_3}$, where ${\bf E}(t)$ is
the incident electric field and $\textit{I}_3$ is the 3x3 identity matrix. In
our calculations, ${\bf E}(t)$ is assumed to be linearly polarized along $x$
axis; thus $\hat{V}(t)=q \, x \, E_x(t)\, \textit{I}_3$ and $\zeta=0$. \par
Substituting Eq.\ref{eq:dipolepsi} into Eq.\ref{eq:dipoletdde}, premultiplying
the result
by $[\psi^{(1)}_{\bf k'}({\bf r})]^\dagger$, and using the real-space
orthogonality property of the eigenstates, we obtain:
\begin{multline}
i\hbar(2\pi)^{2}[\dot{c}_{1}(t)(\frac{k_{y}k_{y}'}{k k'}+\frac{k_{x}k_{x}'}{k k'})+\dot{c}_{2}(t)(\frac{k_{y}k_{x}'}{\sqrt{2}k k'}-\frac{k_{x}k_{y}'}{\sqrt{2}k k'})\\
+\dot{c}_{3}(t)(\frac{k_{y}k_{x}'}{\sqrt{2}k k'}-\frac{k_{x}k_{y}'}{\sqrt{2}k k'})]\delta({\bf k}-{\bf k'})\\
 =eE_{x}(t)\int xe^{i({\bf k}-{\bf k'})\cdot{\bf r}}[c_{1}(t)(\frac{k_{y}k_{y}'}{k k'}+\frac{k_{x}k_{x}'}{k k'})\\
 +c_{2}(t)(\frac{k_{y}k_{x}'}{\sqrt{2}k k'}-\frac{k_{x}k_{y}'}{\sqrt{2}k k'})+c_{3}(t)(\frac{k_{y}k_{x}'}{\sqrt{2}k k'}-\frac{k_{x}k_{y}'}{\sqrt{2}k k'})]d^{2}\bf r
\label{eq:psi1result}
\end{multline}
Integrating Eq. \ref{eq:psi1result} over the $\bf k'$ space and using the
transformations: $c_{1}(t)=\tilde{c}_{1}(t), \ 
c_{2}(t)=\tilde{c}_{2}(t)\exp(i\omega_{p}t), \ \mathrm{and} \ 
c_{3}(t)=\tilde{c}_{3}(t)\exp(-i\omega_{p}t)$ yields the equation of motion for
$\dot{\tilde{c}}_{1}(t)$:
\begin{equation}
\dot{\tilde{c}}_{1}(t)=-\frac{eE_{x}(t)}{\sqrt{2}\hbar k^2}(\tilde{c}_{2}(t)\exp(i\omega_{p}t)+\tilde{c}_{3}(t)\exp(-i\omega_{p}t))k_y.
\label{eq:c1t}
\end{equation}
where $\omega_{p}=v_{0}k$ is the eigenfrequency.


Similarly, following the procedure above with $[\psi^{(2)}_{\bf k'}({\bf r})]^\dagger$ and
$[\psi^{(3)}_{\bf k'}({\bf r})]^\dagger$ yields the following equations of motion for
$\dot{\tilde{c}}_{2}(t)$ and $\dot{\tilde{c}}_{3}(t)$ respectively:
\begin{equation}
\dot{\tilde{c}}_{2}(t)=\frac{eE_{x}(t)}{\sqrt{2}\hbar k^2}\tilde{c}_{1}(t)\exp(-i\omega_{p}t)k_y.
\label{eq:c2t}
\end{equation}

\begin{equation}
\dot{\tilde{c}}_{3}(t)=\frac{eE_{x}(t)}{\sqrt{2}\hbar k^2}\tilde{c}_{1}(t)\exp(i\omega_{p}t)k_y.
\label{eq:c3t}
\end{equation}

\begin{figure}[t]
\includegraphics[width=0.45\textwidth]{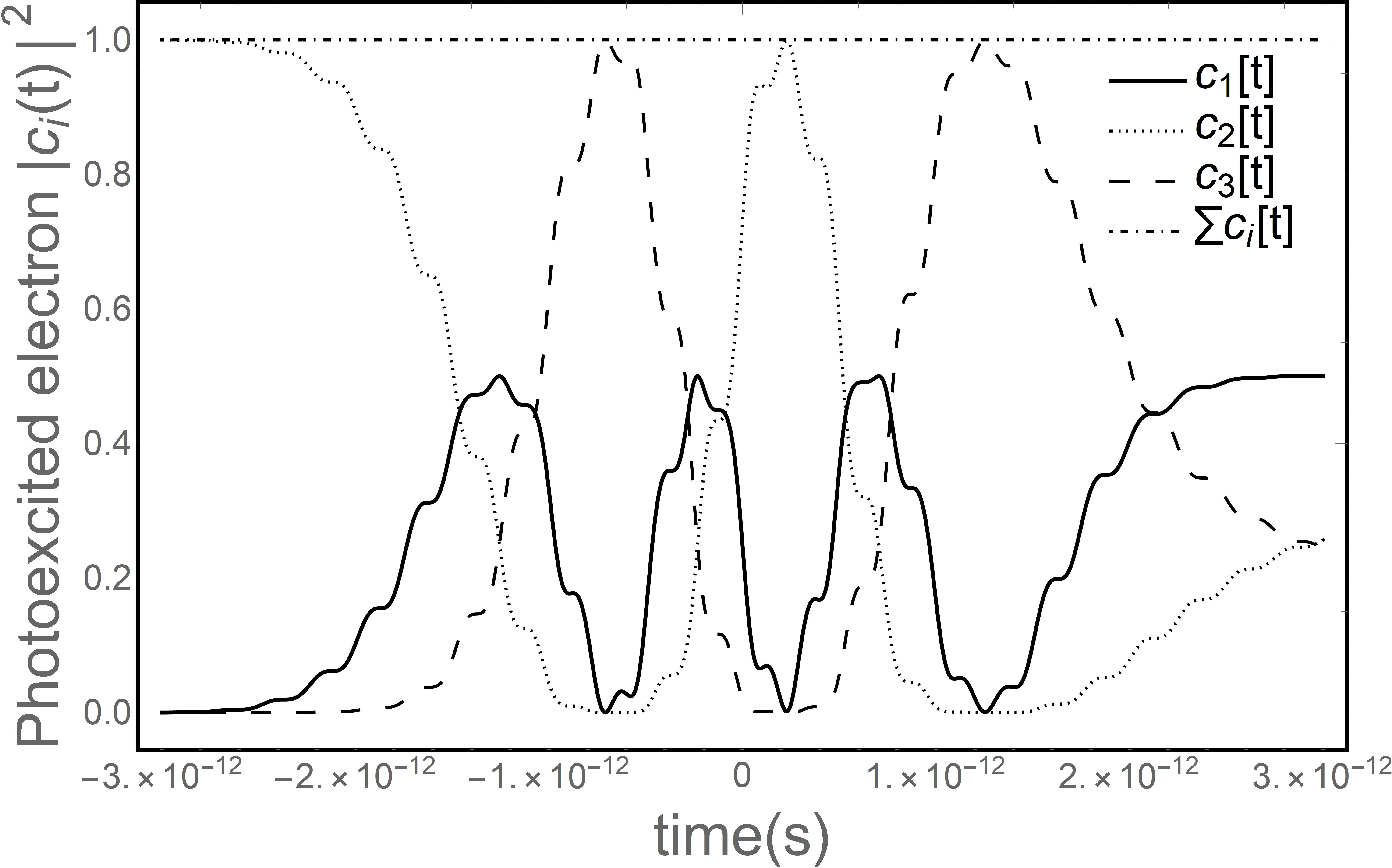}
\caption{\label{fig:epsart} Electron dynamics at a resonant state
$k=\omega_{0}/v_0$ and $\phi_{p}=\pi/4$ with incident light field $f_{0}=2 \,
\mathrm{THz}$,
$\tau=1 \,\mathrm{ps}$, and $I=1 \, \mathrm{kW/cm^2}$ }
\label{fig:rabioscillations}
\end{figure}

Solving the coupled equations Eqs. \ref{eq:c1t}, \ref{eq:c2t}, and \ref{eq:c3t}
numerically using a pulsed THz pump, we can observe the dynamical behavior of
the system, including Rabi oscillations. To gain a detailed understanding of this
behavior, we first consider resonant excitation with $k=\omega_{0}/v_0$,
directional angle of momentum $\phi_{p}=\pi/4$, and light field intensity $I=1
kW/cm^2$. We assume that the electron population is initially in the valence
band. Pump parameters $f_{0}=2$ THz and $\tau=1$ ps are used in our
calculations. Fig. \ref{fig:rabioscillations} shows the temporal evolution of
the normalized band populations $|c_{1}(t)|^2$, $|c_{2}(t)|^2$
and $|c_{3}(t)|^2$ for this configuration. With this parameter set, the Rabi frequency of
the central flat band is nearly double that of the valence and conduction band,
and the Rabi frequencies of valence and conductions are nearly the same. In
addition to the Rabi oscillation, there is a high-frequency oscillation
component observed. This component arises due to the
counter-rotating wave oscillating at $(\omega_{0}+2\omega_p)$.

\subsection{Induced Current Density in LCS Lattice}
In order to simplify the notation used, at this point we transform the
coefficients $\tilde{c}_i (t)$ into population inversions:
$n_{12}(t)=|\tilde{c}_{1}(t)|^2-|\tilde{c}_{2}(t)|^2$,
$n_{31}(t)=|\tilde{c}_{3}(t)|^2-|\tilde{c}_{1}(t)|^2$,
$n_{32}(t)=|\tilde{c}_{3}(t)|^2-|\tilde{c}_{2}(t)|^2$, together with density
matrix elements: $\rho_{ij}=\tilde{c}_{i}{\tilde{c}_{j}}^{\ast}, (i,j=1,2,3)$.
By introducing the continuity equation $\frac{\partial\rho}{\partial
t}+\nabla\cdot{\bf j}=0$ with charge density $\rho=q|\Psi({\bf r},t)|^2$, we
obtain expressions for the single-particle current density $j_{\nu}=e\Psi({\bf r},t)^{\dagger}\frac{\partial \textit{H}}{\hbar \partial k_{\nu}}\Psi({\bf r},t)$, where $\nu=x,y$ indicates the
current density component, as follows:
\begin{subequations}
\begin{equation}
j_x{(t)}=q v_0 (1/2\pi)^{2}[\frac{n_{32}k_x}{k}+\frac{(\rho_{12}+\rho^{\ast}_{12})k_{y}-(\rho_{31}+\rho^{\ast}_{31})k_y}{\sqrt{2}k}],
\end{equation}
\begin{equation}
j_y{(t)}=q v_0 (1/2\pi)^{2}[\frac{n_{32}k_y}{k}+\frac{(\rho_{31}+\rho^{\ast}_{31})k_{x}-(\rho_{12}+\rho^{\ast}_{12})k_x}{\sqrt{2}k}].
\end{equation}
\end{subequations}

Finally, the net (total) current density is obtained by integrating the
single-particle current density over the momentum space:
\begin{equation}
{\bf J}(t)=g_{s}\sum_{\bf k}{\bf j}(t),
\label{eq:totalcurrent}
\end{equation}
where ${g_s}=2$ is the spin-degeneracy factor.

\section{RESULTS AND DISCUSSION\label{sec:resultsanddiscussion}}
\subsection{Single-Particle Harmonic Spectra with Rabi Oscillations}
\begin{figure}[ht]
\begin{subfigure}[h]{0.2\textwidth}
  \centering
  \caption{}
  \includegraphics[width=\textwidth]{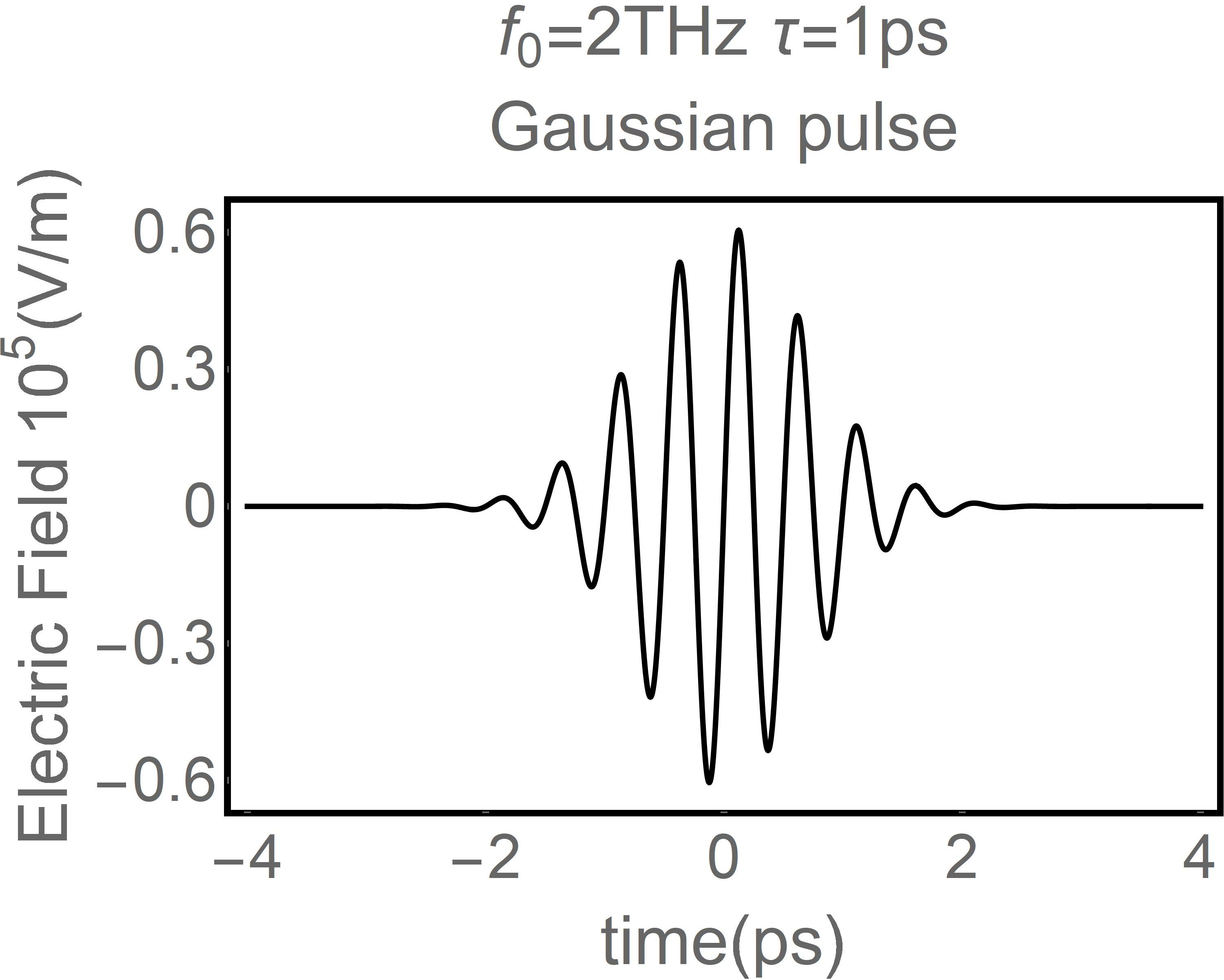}
  \label{fig:rabispectruma}
\end{subfigure}
\quad
\begin{subfigure}[h]{0.2\textwidth}
  \centering
  \caption{}
  \includegraphics[width=\textwidth]{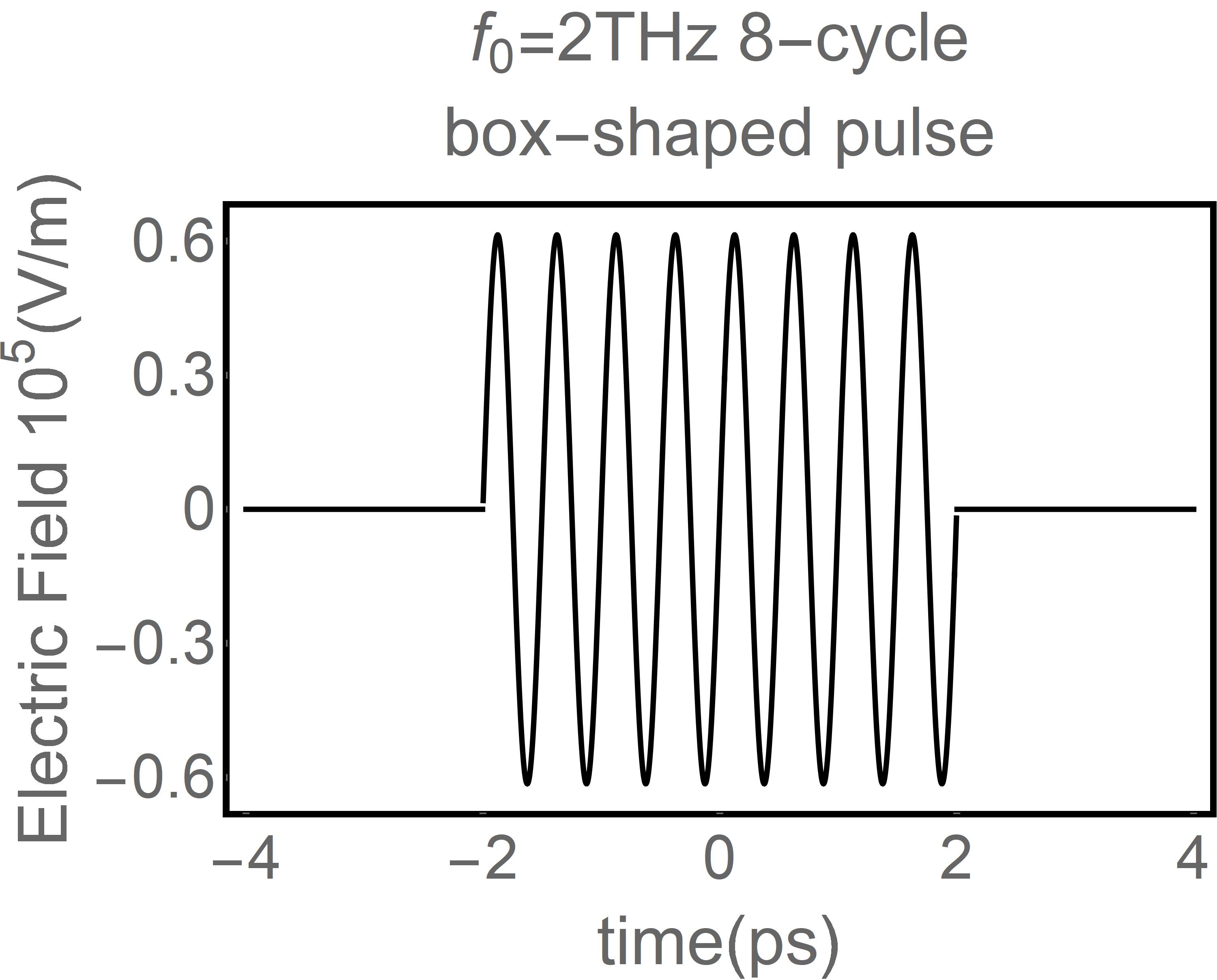}
  \label{fig:rabispectrumb}
\end{subfigure}\\
\begin{subfigure}[h]{0.2\textwidth}
  \centering
  \caption{}
  \includegraphics[width=\textwidth]{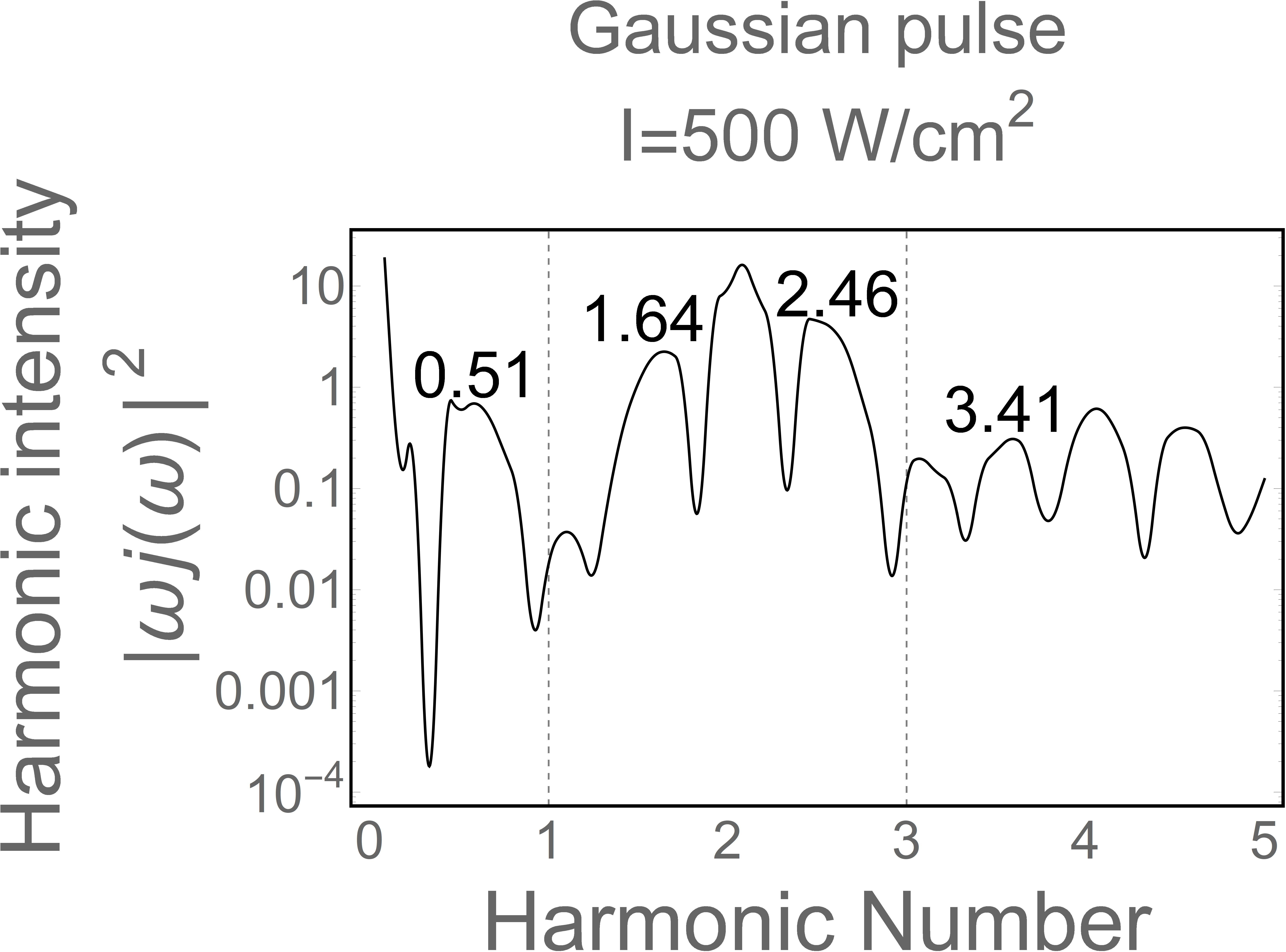}
  \label{fig:rabispectrumc}
\end{subfigure}
\quad
\begin{subfigure}[h]{0.2\textwidth}
  \centering
  \caption{}
  \includegraphics[width=\textwidth]{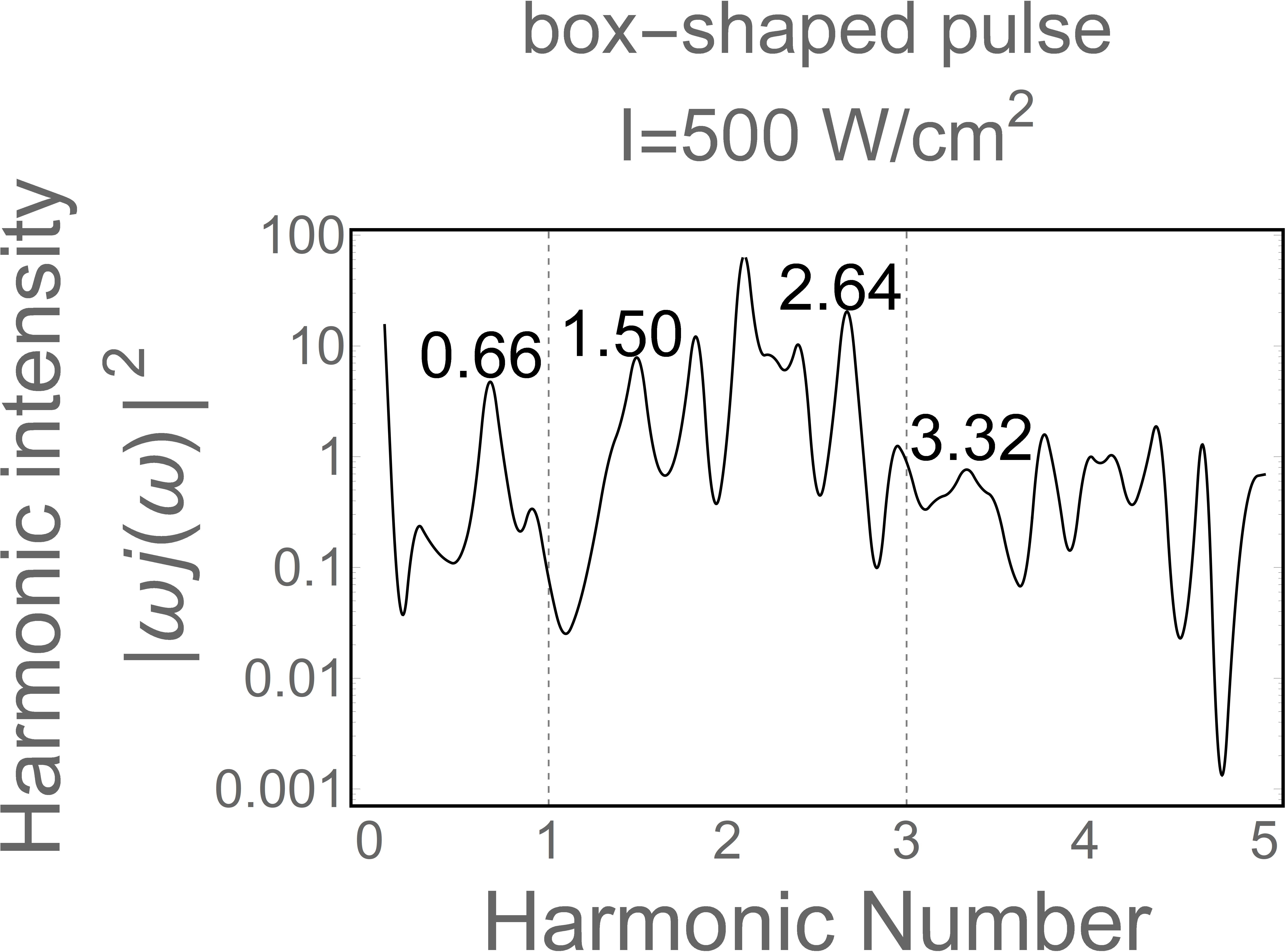}
  \label{fig:rabispectrumd}
\end{subfigure}
\caption{Illustration of the contribution from Rabi oscillations to the
single-particle current
spectra at $k=\omega_{0}/v_0$ with $\phi_{p}=\pi/4$ in the LCS
lattice. A Gaussian pulse (a) with peak irradiance (c) $I=500W/cm^2$. A
box-shaped pulse (b) with peak irradiance (d) $I=500 \,
\mathrm{W/cm^2}$.}
\label{fig:rabispectra}
\end{figure}

In Fig. \ref{fig:rabispectra}, we plot the single-particle current spectra
excited by a Gaussian (Fig. \ref{fig:rabispectruma}) or square (Fig.
\ref{fig:rabispectrumb}) pulse for resonant excitation at $k=\omega_0/\nu_0$ and
$\phi_k = \pi/4$ with pump irradiance $I = 500 \, \mathrm{W/cm^2}$. The
resultant single-particle current spectra are plotted in
Figs. \ref{fig:rabispectrumc} and \ref{fig:rabispectrumd} respectively.
These spectra exhibit energy at both even and odd
harmonics. Further, the bandwidth over which significant energy content exists
is quite large, ranging over at least the lowest five harmonics of the 2 THz
fundamental excitation frequency. The detailed pulse shape is shown to impact
the spectrum in a manner that is to be expected. The square pulse gives rise to
sharper peaks in the spectrum than does the Gaussian pulse.

\begin{figure}[htp]
\begin{subfigure}[h]{0.2\textwidth}
  \centering
  \caption{}
  \includegraphics[width=\textwidth]{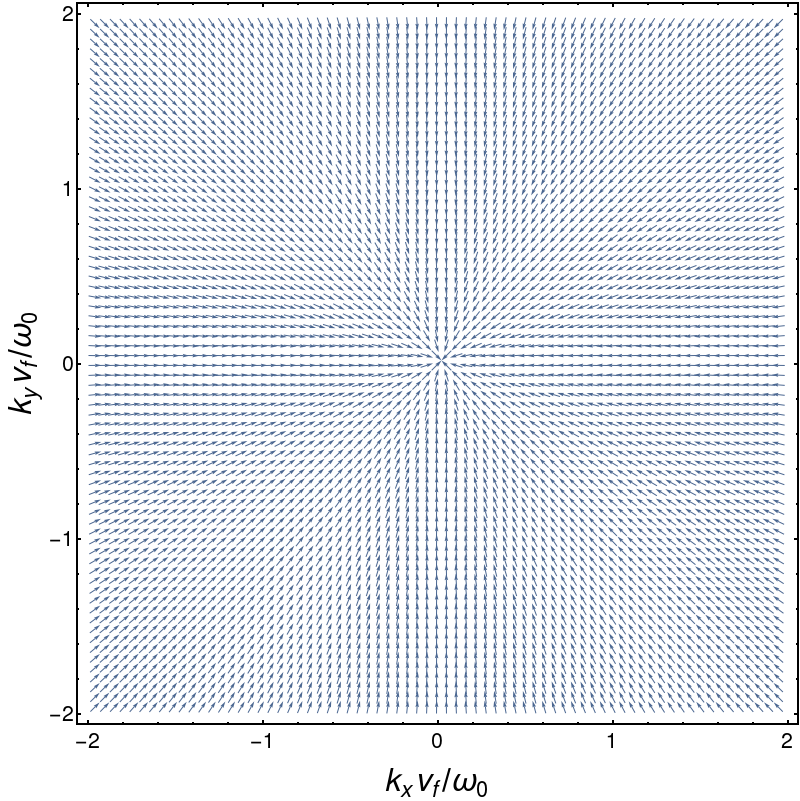}
  \label{fig:000delay}
\end{subfigure}
\quad
\begin{subfigure}[h]{0.2\textwidth}
  \centering
  \caption{}
  \includegraphics[width=\textwidth]{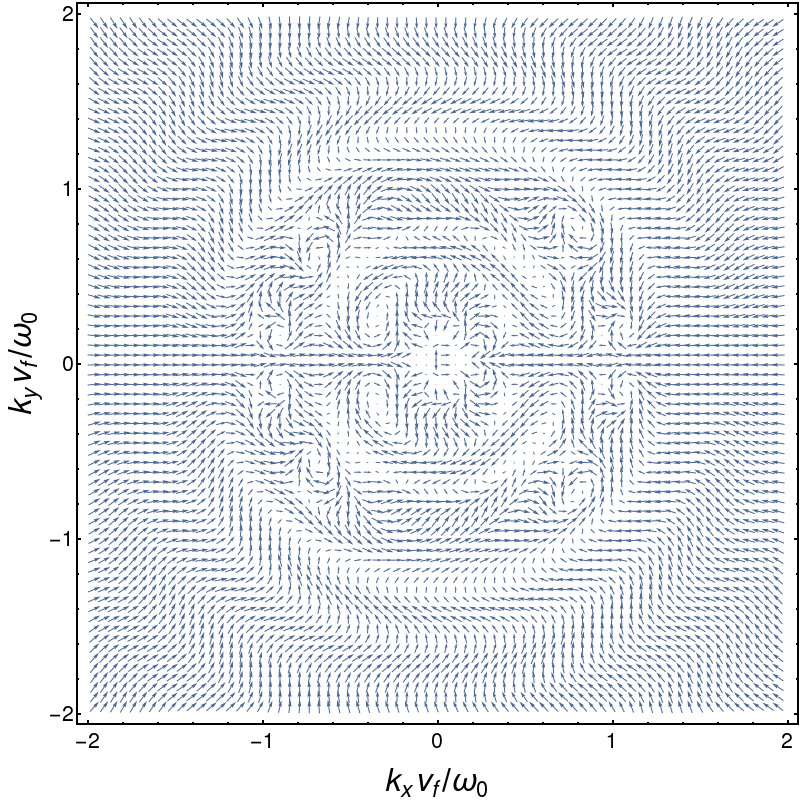}
  \label{fig:220delay}
\end{subfigure}\vspace{-0.175in}\\
\begin{subfigure}[h]{0.2\textwidth}
  \centering
  \caption{}
  \includegraphics[width=\textwidth]{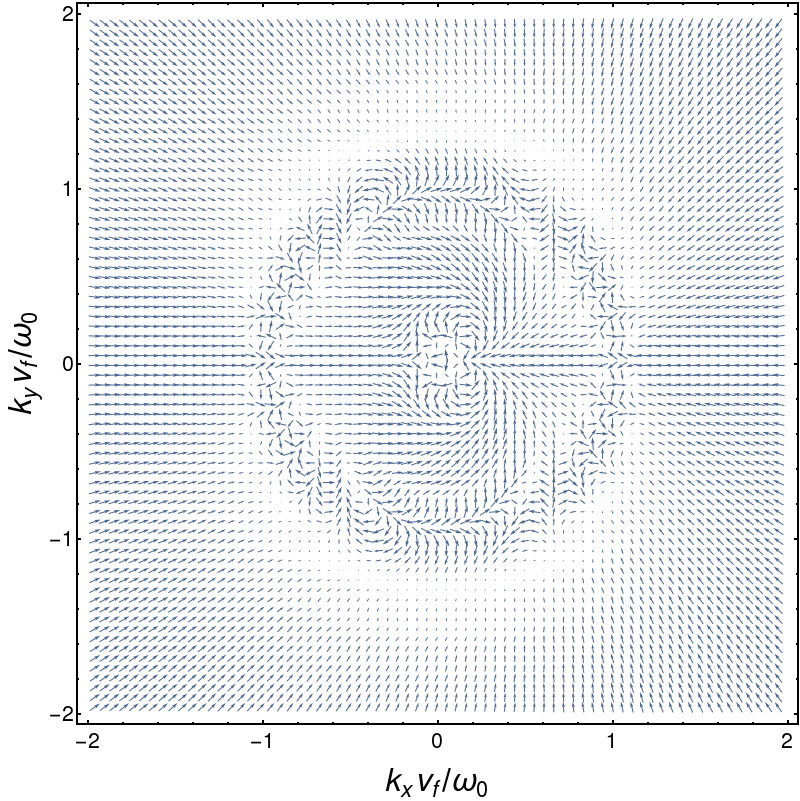}
  \label{fig:298delay}
\end{subfigure}
\quad
\begin{subfigure}[h]{0.2\textwidth}
  \centering
  \caption{}
  \includegraphics[width=\textwidth]{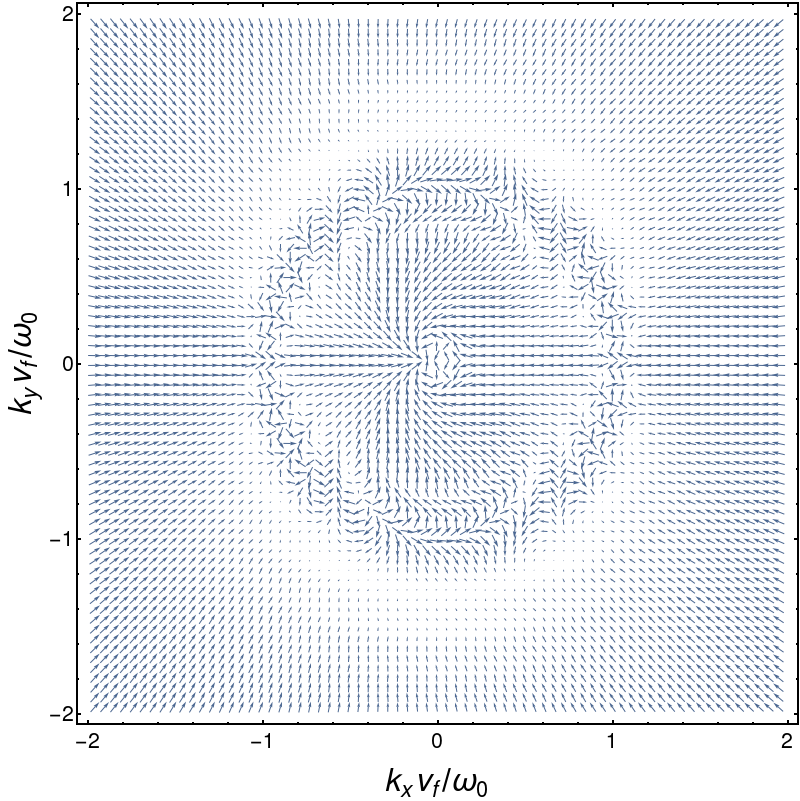}
  \label{fig:301delay}
\end{subfigure}\vspace{-0.175in}\\
\begin{subfigure}[h]{0.42\textwidth}
  \centering
  \caption{}
  \includegraphics[width=\textwidth]{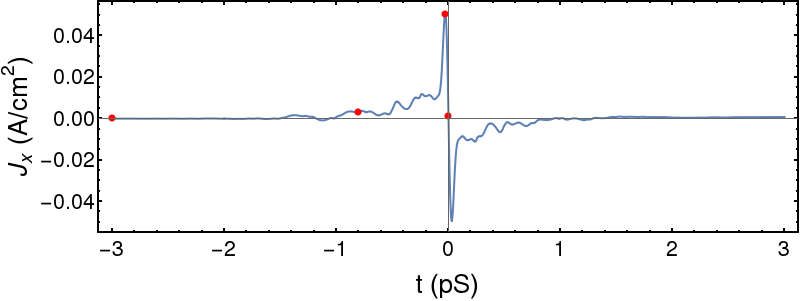}
  \label{fig:totaltemporalcurrent}
\end{subfigure}\vspace{-0.175in}\\
\begin{subfigure}[h]{0.42\textwidth}
  \centering
  \caption{}
  \includegraphics[width=\textwidth]{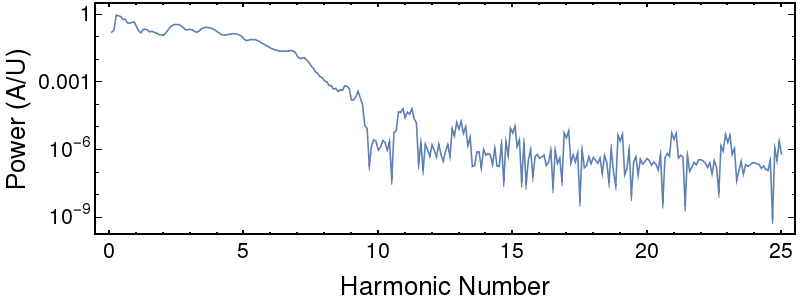}
  \label{fig:totalspectrumcurrent}
\end{subfigure}\\
\caption{Current density for a peak pump irradiance of $I = 10 \,
\mathrm{kW/cm^2}$. Plots a) through d) illustrate the magnitude and direction of
the single-particle current density $\mathbf{j}$ at increasing time samples
throughout the pump pulse. Plot e) shows the total current density $\mathbf{J}$ with
the red dots indicating the temporal position of the single particle current
density frames a)-d). Finally, plot f) shows the spectral power density of the total
current density $\mathbf{J}$.}
\label{fig:totalcurrentspectrum}
\end{figure}

\subsection{Total Current Spectral Content}
Following Eq. \ref{eq:totalcurrent}, we obtain the total current by integrating
numerically over $\mathbf{k}$. We have performed this calculation for pump
irradiances ranging from $I = 100 \, \mathrm{W/cm^2}$ to $I = 10 \, \mathrm{kW/cm^2}$.
In what follows, we characterize the spectral content of the total current over
this complete pump range, however in the interest of conserving space, we plot
the results only for the $I = 10 \, \mathrm{kW/cm^2}$ case in Fig.
\ref{fig:totalcurrentspectrum}.

In Figs. \ref{fig:000delay}-\ref{fig:301delay}, we illustrate the magnitude
and direction of the single-particle current density over a range of times relative to
the pump pulse. Fig. \ref{fig:000delay} illustrates the thermal equilibrium
current distribution prior to the arrival of the pump pulse. Figs.
\ref{fig:220delay}, \ref{fig:298delay}, and \ref{fig:301delay} show the
single-particle current density for times $800 \, \mathrm{fs}$, $6 \,
\mathrm{fs}$, and $0 \, \mathrm{fs}$ prior to
the peak irradiance of the pump pulse respectively. These arrival times are noted as red
dots on Fig. \ref{fig:totaltemporalcurrent}, which plots the temporal evolution
of the total current density $\mathbf{J}$.
We note that in our model, the total current density does not decay exactly to
zero as the current dynamics disappear due to the passing pump pulse. Such a
result is a consequence the fact that our model does not include a
relaxation term. The absence of a relaxation term, coupled with the fact that the area of
the $I = 10 \, \mathrm{kW/cm^2}$ pulse does not return the system exactly to its
initial condition, results in a persistent offset in the plot of the temporal
current density. Such an offset does not materially affect our conclusions.

Finally, in Fig. \ref{fig:totalspectrumcurrent}, we plot the spectrum of the
total current density $\mathbf{J}$. The spectrum exhibits a continuum that
persists up through the ninth harmonic. This continuum rolls off by two orders
of magnitude at the high-frequency cutoff. At higher frequencies, the energy is
localized around the odd harmonics as would be expected from a perturbation
analysis of the problem due to the symmetry inherent in the LCS lattice. Each of
the odd harmonics exhibits a bifurcation into a higher and lower frequency lobe
surrounding the odd harmonic.
This large spectral continuum may be understood by considering the evolution
equations, Eqs. \ref{eq:c1t}, \ref{eq:c2t}, and \ref{eq:c3t}. In this set of
equations, the Rabi frequencies are proportional to $E_x(t) \sin\theta_k/k$.
As a result, the frequency content of the total current ranges from a
component with infinite frequency at $k=0$ to components with frequency
asymptotically approaching 0 as $k \rightarrow \infty$. These components are
weighted proportionally to $2 \pi k$ due to the increasing density of states as
$k$ increases.

Examining the resultant spectral content of the total current density for lower
peak pump irradiances, we observe the following:
for a peak pump irradiance of $1 \, \mathrm{kW/cm^2}$, the continuum spectrum
is down by two orders of magnitude relative to the low-frequency component of
the total current density. Harmonics 1 (fundamental) and 3 bifurcate, whereas harmonics 5-11
are visible but do not bifurcate. For a peak pump irradiance of $100 \,
\mathrm{W/cm^2}$, the continuum is below the low frequency components by
approximately the same two orders of magnitude, however only the fundamental
frequency component bifurcates. Frequency components at harmonics 3-9 exist, but
do not bifurcate at this irradiance.

Finally, we note that due to the mirror symmetry of the single-particle current
density in the direction normal to the applied pump polarization, the total
current in the direction normal to the polarization is zero.

\section{Conclusion\label{sec:conclusion}}
In conclusion, we have analyzed the three-level LCS lattice and obtained
closed-form expressions for the carrier dynamics in this system under the
influence of a picosecond THz pump pulse polarized in the plane of the lattice.
The total current density $\mathrm{J}$ arising in this system exhibits a
spectral content that evolves toward a continuum at
relatively moderate pump irradiances. We provide a detailed analysis of that
continuum for a pump irradiance of $10 \, \mathrm{kW/cm^2}$.
\begin{acknowledgments}
Q. Jin acknowledges partial support from a U of Iowa undergraduate research
fellowship.
\end{acknowledgments}

\nocite{*}
\bibliography{refs}

\end{document}